\documentstyle[12pt,epsf]{article}
\large

\def\ga{\mathrel{\mathpalette\fun >}}
\def\fun#1#2{\lower3.6pt\vbox{\baselineskip0pt\lineskip.9pt
\ialign{$\mathsurround=0pt#1\hfil##\hfil$\crcr#2\crcr\sim\crcr}}}

\newcommand{\bc}{\begin{center}}
\newcommand{\ec}{\end{center}}
\newcommand{\bd}{\begin{displaymath}}
\newcommand{\ed}{\end{displaymath}}
\newcommand{\be}{\begin{equation}}
\newcommand{\ee}{\end{equation}}
\newcommand{\ba}{\begin{array}}
\newcommand{\ea}{\end{array}}
\newcommand{\bt}{\begin{tabular}}
\newcommand{\et}{\end{tabular}}

\begin{document}

\begin{center}
\large
THE STRUCTURE FUNCTIONS OF THE  \\
PHOTON AT LOW VIRTUALITIES
\end{center}

\begin{center}
 {\bf I.A. Shushpanov}
\end{center}
\begin{center}
\it{Institute for Theoretical and Experimental Physics, B.
Cheremushkinskaya 25, Moscow 117218, Russia}
\end{center}

\vspace{1.cm} \centerline{January 2002} \vspace{1.5cm}
\begin{abstract}

The structure functions $F_1$ and $F_2$ of the real photon at low
virtualities are calculated in the framework of chiral pertubation
theory(ChPT) in the zero and first order of ChPT. It is assumed
that the virtuality of hard projectile photon $Q^2$ is much less
than the characteristic ChPT scale. In this approximation the
structure
functions are determined by the production of two pions in $\gamma
\gamma$ collisions. The numerical results for $F_2$ and $F_1$ are
presented.
\end{abstract}

\section{ Introduction}
The chiral perturbation theory (ChPT) is the convenient tool for
obtaining quantitative predictions in the low-energy region
\cite{GL1,GL2}. The properties of the photon at low virtualities
also can be described by this theory. The cross section for
$\gamma\gamma\rightarrow\pi^+ \pi^-$ has been calculated some
years ago \cite{bj}. It is possible to extend this method to the
case when both target and projectile photons are virtual. For
example, the structure functions of longitudinal virtual photon
were calculated in [Me]. In this paper we perform the calculations
of the real photon structure functions at low $Q^2$ basing on
ChPT. In order to be in the framework of ChPT it is necessary to
assume that virtuality of probe photon $Q^2$ is much less than the
applicability limit of ChPT $\Lambda^2 \approx m_{\rho}^2 =
0.6$GeV$^2$. For the value of center of mass energy $s=(p+q)^2$
the more weak limitation is imposed: $s \sim \Lambda^2$ because
$s$ explicitly enters only in the small corrections to the
results. It means that the photon structure functions are not in
the scaling region of $Q^2$ and our results can not be directly
compared with another theoretical predictions for structure
functions of real photon at large $Q^2$ \cite{iof}. However, it
seems useful to have the data from the both sides of region of
intermediate $Q^2$.

\section{Calculation of the Structure Functions}

In order to get the structure functions $F_2$ and $F_1$ of the
real target  photon the imaginary part of $\gamma \gamma$ forward
scattering amplitude ${\rm Im} T_{\mu\nu\lambda\sigma}(p,q)$ must
be multiplied by the product of the photon polarization vectors in
the initial and final states
\be
\frac{1}{2\pi} \Sigma e_\lambda
e_\sigma^* {\rm Im } T_{\mu\nu\lambda\sigma}(p,q)=(-g_{\mu\nu}
+q_\mu q_\nu/q^2)F_1+ (p_\mu -q_\mu \nu/q^2)(p_\nu -q_\nu
\nu/q^2)F_2/\nu,
\ee
where $\nu=pq$. By the standard definition of
the structure functions the factor $e^2$ arising from
electromagnetic vertices with projectile photon should be omitted
in the amplitude. For the case of real unpolarized photon  the
photon density matrix $\Sigma e_\lambda e_\sigma^*$ can be written
as $-g_{\lambda\sigma}/2$.

In the zero order of ChPT the chiral Lagrangian reduces to the
Lagrangian of scalar electrodynamics
\be
\label{scal}
L^{scal.elec.}=(\partial_\mu \pi+iA_\mu \pi)^+ (\partial_\mu
\pi+iA_\mu \pi)- M_\pi^2 \pi^+ \pi
\ee
and $\gamma \gamma$ forward
scattering amplitude is determined by the two box diagrams in
Fig.1 and the diagrams of Fig. 2, where two photon interact with
two pions at one point - the diagrams arising from
$A_\mu^2\pi^+\pi$ term in Lagrangian of scalar electrodynamics.

\begin{figure}
\centerline{\epsfbox{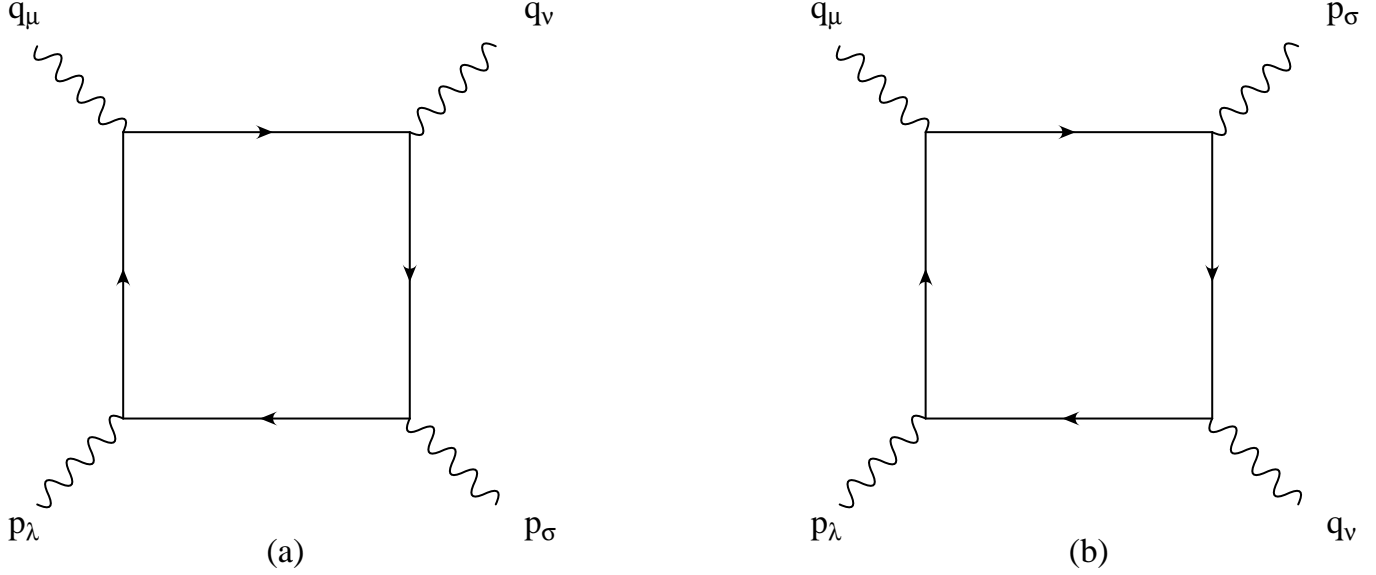}}
\caption{The forward $\gamma\gamma$ scattering amplitude box diagrams
 in scalar electrodynamics -- the zero order of ChPT; the solid lines
 correspond to pions, the wavy lines to photons, a) direct diagram;
b) crossing diagram.}
\end{figure}

\begin{figure}
\centerline{\epsfbox{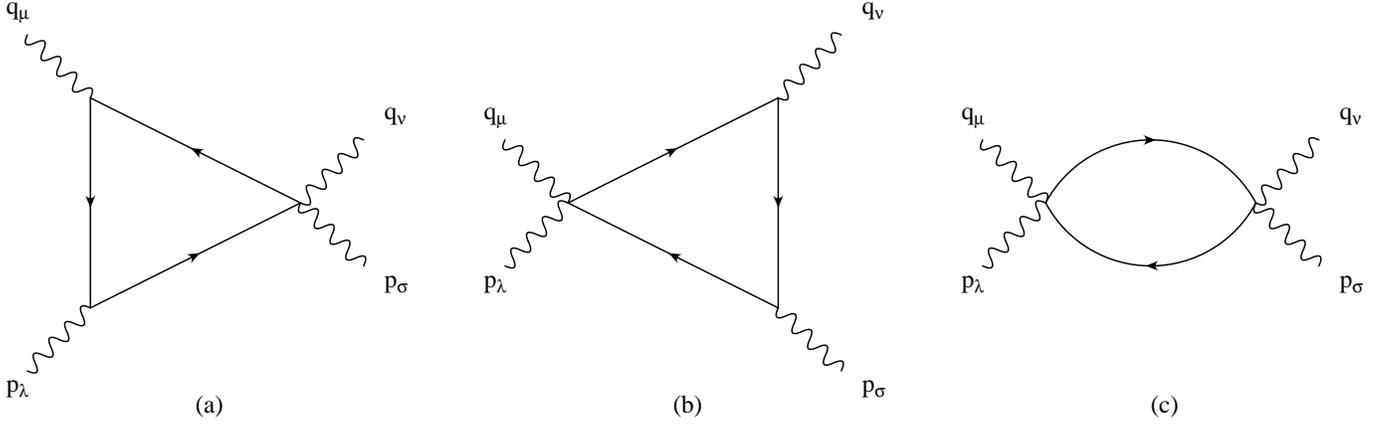}}
\caption{The diagrams of scalar electrodynamics for the  forward
$\gamma\gamma$ scattering, arising from the term $A_{\mu}^2\pi^+\pi$
in the Lagrangian.}
\end{figure}

The imaginary part of the amplitude is connected with discontinuity of
the amplitude in $s$-channel by:

\be
{\rm Im}T_{\mu\nu\lambda\sigma}(p^2,q^2,s)=\frac{1}{2i}
[T_{\mu\nu\lambda\sigma}(p^2,q^2,s+i0)
-T_{\mu\nu\lambda\sigma}(p^2,q^2,s-i0)].
\ee

It is convenient to calculate $F_2$ and $F_1$
structure functions as the coefficient before $p_\mu p_\nu$
and $-g_{\mu\nu}$, respectively. Doing in this way we find the
contributions of diagrams in Fig.1,2  to the structure
functions (at $p^2=0$): for the direct diagram
\be
\label{F21a}
F^{1a}_2=\frac{\alpha}{4\pi}
\left\{-2x^2 L +(1+2x-10x^2+4x^3)\phi +
\frac{2M_\pi^2}{\nu}[x(5-6x)L+(-1+8x-8x^2)\phi]
\right\}
\ee
\be
\label{F11a}
F^{1a}_1=\frac{\alpha}{4\pi}
\left\{(1-x)\phi +
\frac{M_\pi^2}{\nu}[(3-2x)L+4(1-x)\phi]
\right\},
\ee
for the crossing diagram
\be
\label{F21b}
F^{1b}_2=\frac{\alpha}{4\pi}
\left\{(1-2x)[2x(1-x)L +(6x-6x^2-1)\phi] +
\frac{2M_\pi^2}{\nu}[-x L+(1-6x+6x^2)\phi-2x\frac{M_\pi^2}{\nu}L]
\right\}
\ee
\be
\label{F11b}
F^{1b}_1=\frac{\alpha}{4\pi}
\left\{(1-3x+2x^2)\phi +
\frac{M_\pi^2}{\nu}[(1-2x)L-2(1-x)\phi-\frac{2M_\pi^2}{\nu}L],
\right\}
\ee
and for the contact diagrams
\be
\label{F22ab}
F^{2a,b}_2=\frac{\alpha}{2\pi}
\left\{x(2x-1)L +(-4x+6x^2)\phi -
\frac{2M_\pi^2}{\nu}x L
\right\};
\qquad F^{2c}_2=0
\ee
\be
\label{F12ab}
F^{2a,b}_1=\frac{\alpha}{2\pi}
\left\{(-1+x)\phi - \frac{M_\pi^2}{\nu}L
\right\};
\qquad F^{2c}_1=\frac{\alpha\phi}{4\pi}
\ee

Here $x=Q^2/2\nu$ is the Bjorken variable, $\phi$ accounts two pion
phase space,
\be
\phi=\left(1-\frac{4M_\pi^2 x}{Q^2 (1-x)}\right)^{1/2}, \qquad
L=\ln\left[\frac{1-\phi}{1+\phi}\right].
\ee

The next step which can be done is the calculation of ChPT corrections
induced by pion self-interaction.
 The general method of calculation of these corrections is exposed in
\cite{GL1,GL2}. From the general form of ChPT effective Lagrangian it can be easily
shown that in the first order in ChPT only the two pions intermediate
states can contribute to the imaginary part of the forward
$\gamma\gamma$-scattering amplitude.

 To calculate corrections to the structure functions $F_2$ and $F_1$
in the first order of ChPT one should consider the general expression for
effective Lagrangian which contains all terms permitted by chiral invariance
up to the order (momentum)$^4$.

In the leading order of ChPT the effective Lagrangian has the form
\be
\label{ZO}
L=\frac{F_\pi^2}{4} {\rm Tr} \{\nabla_\mu U\nabla_\mu U^\dagger\}+
\Sigma {\rm Re\ Tr} \{{\cal M}U^\dagger\} + \quad \cdots
\ee
where $U$ is the unitary matrix, $U=U^0 +i U^i \tau^i$ and
$U^i=\pi^i/F_\pi$ in Weinberg parameterization. $F_\pi=93$MeV is
the pion decay constant and parameter $\Sigma$ has the meaning
of quark condensate.
The ChPT expansion corresponds
to the expansion in inverse powers of $F_\pi^2$. So, in the first
order of ChPT
\be
U^0=1-\frac{1}{2}U^i U^i
\ee
The covariant derivative $\nabla_{\mu} U$ is defined by
\be
\nabla_\mu U=\partial_\mu U + i[U,V_\mu]
\ee
where $V_{\mu}=A_{\mu} \tau^3/2$ is the electromagnetic field.
Expressing the real
pion fields $\pi^1$ and $\pi^2$ through the fields of charged pions
$\pi^+$ and $\pi$
\be
\pi^1=\frac{1}{\sqrt{2}}(\pi+\pi^+)\quad
\pi^2=\frac{1}{\sqrt{2}i}(\pi-\pi^+)
\ee
and expanding ($\ref{ZO}$) up to terms $\sim1/F_\pi^2$, we have (only charged
pion fields are retained)
\be
L^{(2)}=L^{scal.elec.}+\frac{1}{2F_\pi^2}\{\partial_\mu(\pi^+\pi)\}^2-
\frac{M_\pi^2}{2
F^2_\pi} [\pi^+ \pi^- ]^2,
\ee
where
$L^{scal.elec.}$ defined in ($\ref{scal}$). Second and third terms in $L^{(2)}$ corresponds to the four pion
interaction and leads to appearance of the loop corrections to the structure
functions which are proportional to $1/F_\pi^2$.

 In the order (momentum)$^4$, i.e. $\sim1/F_\pi^2$, there are two terms
in the general effective Lagrangian of ChPT \cite{GL1,GL2},
which are essential for us and they can be written in terms of
charged pion fields
\be
L^{(4)}=-\frac{2 l_5}{F_\pi^2} (e F_{\mu\nu})^2 \pi^+ \pi^- -
\frac{2i l_6}{F_\pi^2} e F_{\mu\nu} [\partial_\mu \pi^-
\partial_\nu \pi^+
+ ie A_\mu \partial_\nu (\pi^+ \pi^- ) ],
\label{l5l6}
\ee
where $F_{\mu\nu}$ is the electromagnetic field strength and
$l_5$, $l_6$ are some phenomenological
(infinite) constants.

The terms in the Lagrangian $L^{(4)}$ serve as the counter terms
for the renormalization of loops: the infinities arising in loop
calculations are absorbed in $l_5$ and $l_6$, and as a result the
finite values $\bar{l_5}$ and $\bar{l_6}$ arise.

For the calculation of the first order
ChPT corrections to the structure functions we should calculate
effective $\gamma\pi\pi$ and $\gamma\gamma\pi\pi$ vertices,
substitute ones in all zero order diagrams of ChPT and collect terms
proportional to $1/F_\pi^2$.

 At first let us consider effective $\gamma\pi\pi$ vertex. In the chosen
parameterization of U there are three diagrams which contribute to this
vertex in the first order of ChPT (Fig.3).
\begin{figure}
\centerline{\epsfbox{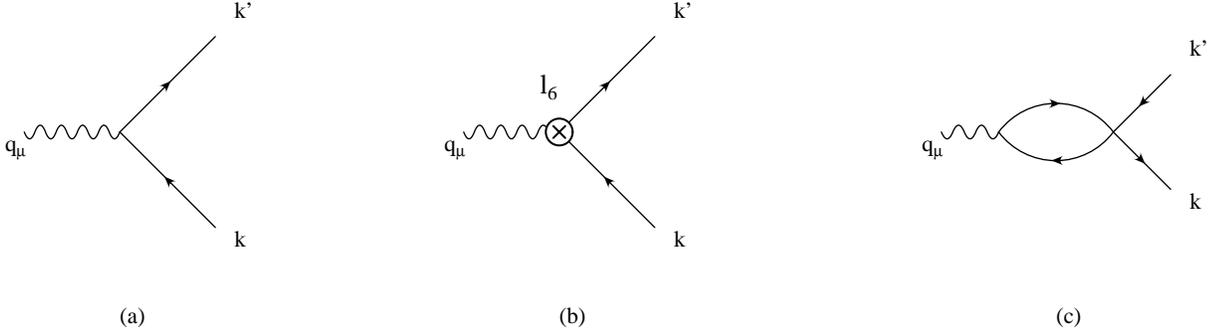}}
\caption{The diagrams for effective $\gamma\pi\pi$ vertex, a) the diagram
corresponding to scalar electrodynamics; b) the diagram arising from
$L_6$ term in chiral Lagrangian; c) the loop diagram corresponding to
four pion interaction.}
\end{figure}

First diagram corresponds
to vertex of $\gamma\pi\pi$ interaction in the scalar electrodynamics,
the second comes from $L_6$ and the third is the loop (unitary) diagram.
 To renormalize the loop
diagram it is convenient to use the dimensional regularization.
 The contribution
of this diagram to the $\gamma\gamma\pi$ vertex can be divided into
finite part and divergent part which contains the infinite factor
$\lambda$, where $\lambda$ is :
$$\lambda=\frac{2}{4-d} + \ln 4\pi+1-\gamma_E -
\ln \frac{M_\pi^2}{\mu^2},$$
and $\mu$ is the scale of mass introduced by dimensional regularization.

 In the sum of the diagrams the divergent part will be absorbed into
$l_6$ and the following result for effective $\gamma\pi\pi$
vertex in the first order of ChPT was obtained \cite{GL1,GL2}:
\be
-\frac{1}{i}\Gamma_\mu(k,k';q)=(k'+k)_\mu-\frac{\left(\bar{l}_6-1/3+
\sigma^2\{\sigma \ln(\frac{\sigma-1}{\sigma+1})+2\}\right)}{48\pi^2F_\pi^2}
(q_\mu kq-k_\mu q^2),
\ee
where
\be
\sigma=(1+4M_\pi^2/Q^2)^{1/2},
\ee
$k$ and $k'$ are the pion initial and final momenta, $q$ is the photon
momentum, $k=k'+q.$ The numerical value of $\bar{l_6}$ was found in
\cite{GL1}
from the data on electromagnetic charge radius of the pion:
\be
\bar{l_6}=16.5\pm 1.1
\ee
  After
substituting this effective vertex in zero order diagrams in all cases
the following combination appears

\be
\label{R6}
R_6(Q^2)=\bar{l}_6-1/3+\sigma^2\left\{\sigma
\ln\left[\frac{\sigma-1}{\sigma+1}\right]+2\right\}.
\ee
The term proportional to $\sigma^2$ in ($\ref{R6}$)
arises from the loop correction -- the diagram Fig. 3c. Numerically,
it is much
smaller numerically (about 10 times) than $\bar{l_6}$.
 In the case of $\gamma\gamma\pi\pi$ effective vertex there are six
diagrams (Fig. 4): first diagram comes from scalar electrodynamics,
second and third ones correspond to $L_6$ and $L_5$, respectively, and
the other diagrams are loop (unitary) corrections.
\begin{figure}
\centerline{\epsfbox{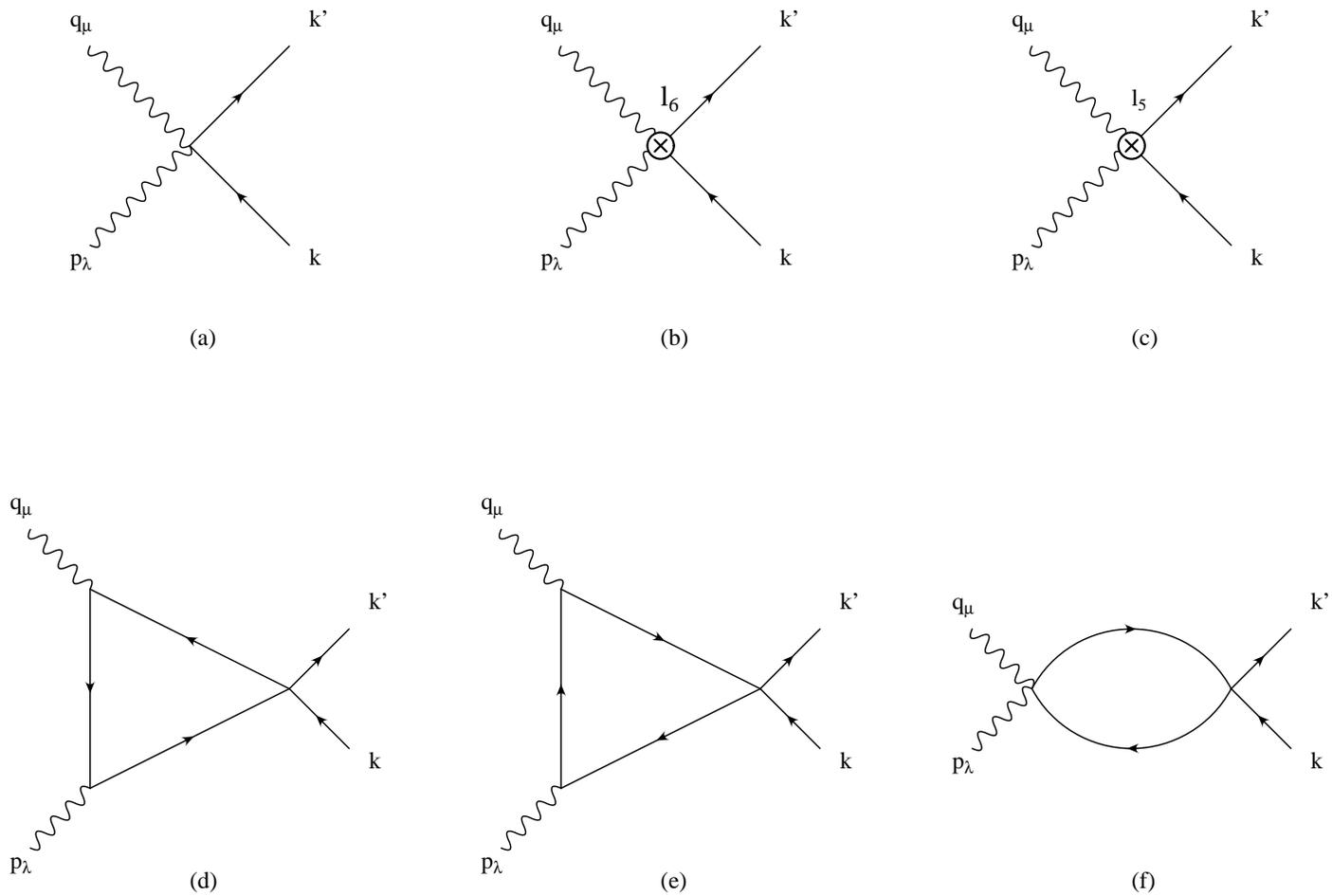}}
\caption{The diagrams for effective $\gamma\gamma\pi\pi$ vertex,
a) the diagram coming from scalar electrodynamics; b), c) the diagrams
corresponding to $L_6$ and $L_5$ terms in chiral Lagrangian,
respectively; d), e), f) the loop diagrams.}
\end{figure}

The calculation of these diagrams leads to the following expression
of  $\gamma\gamma\pi\pi$ effective vertex
$$
\frac{1}{i}\Gamma_{\mu\lambda}(p,q)=2g_{\mu\lambda}-
\frac{(\bar{l_6}-\bar{l_5})}{24\pi^2 F_\pi^2}
(p_\mu q_\lambda -\nu g_{\mu\lambda})
-\frac{s}{F_\pi^2}B_{\mu\lambda}-
$$
\be
-\frac{R_6}{48\pi^2 F_\pi^2}(q_\mu q_\lambda -g_{\mu\lambda}q^2+
p_\mu p_\lambda -g_{\mu\lambda}p^2),
\ee
where $\bar l_6 - \bar l_5 \approx 2.7$ \cite{GL1} and
$$
B_{\mu\lambda}=\frac{1}{i}\int \frac{d^4 k}{(2\pi)^4}\left\{
\frac{(2k+q)_\mu (2k-p)_\lambda}
{[k^2-M_\pi^2][(k-p)^2-M_\pi^2] [(k+q)^2-M_\pi^2]}-\right.
$$
\be
\left.-\frac{g_{\mu\lambda}}
{[(k-p)^2-M_\pi^2] [(k+q)^2-M_\pi^2]}
\right\}.
\ee

In principle we can write out the tensor structure of $B_{\mu\lambda}$
at once:
\be
\label{B}
B_{\mu\lambda}=\frac{\delta}{24\pi^2 s}
(p_\mu q_\lambda -\nu g_{\mu\lambda})+
A(q_\mu -p_\mu q^2/\nu)(p_\lambda -q_\lambda p^2/\nu).
\ee

 This equation follows from the fact that $B_{\mu\lambda}$
 is transverse in $q_\mu$ and $p_\lambda$
simultaneously and can be checked by the direct calculation.

After putting $p^2=0$ second term in ($\ref{B}$) becomes proportional
to the momentum of the target photon $p_\lambda$. Averaging over photon density
matrix actually means contracting of two indices ($\sigma$ and
$\lambda$) in the amplitude
${\rm Im T}_{\mu\nu\lambda\sigma}(p,q)$ corresponding to the target real
photon and contribution of the second term in ($\ref{B}$) to the structure
functions vanishes due to the transversality of
${\rm Im T}_{\mu\nu\lambda\sigma}$. It means that we should find
only $\delta$ which can be calculated by the standard technique
$$
\delta (p^2=0)=3(1-x)\left [2\int_0^1 dy \int_0^y dz \ln
\frac{M_\pi^2 +Q^2 z(1-z)-2\nu z(1-y)}{M_\pi^2}-\right.
$$
\be
\left.-\int_0^1 dy \ln \frac{M_\pi^2 -s y(1-y)}{M_\pi^2}
\right]
\ee

 Substituting these effective vertices into all zero order diagrams
and collecting the terms proportional to $1/F_\pi^2$ we get the final results
for ChPT corrections to structure functions $F_2$ and
$F_1$ of the real photon

\be
F^{ChPT}_2=-\frac{R_6 Q^2}{48\pi^2F_\pi^2}[F^{(1a)}_2 +F^{(1b)}_2
+F^{(2a,b)}_2 /2]
+\frac{\alpha Q^2(\bar l_6 - \bar l_5+\delta)}{96\pi^3F_\pi^2 }
\left(\phi x -\frac{M^2}{\nu}L\right)
\ee
and
\be
 F^{ChPT}_1=-\frac{R_6 Q^2}{48\pi^2F_\pi^2}[F^{(1a)}_1 +F^{(1b)}_1
+F^{(2a,b)}_1 /2]
+\frac{\alpha Q^2[(\bar l_6 - \bar l_5+\delta)-R_6 x]}{192\pi^3F_\pi^2 x}
\left(\phi x -\frac{M^2}{\nu}L\right),
\ee
where $F^{(1a,b,2a,b)}_2$ and $F^{(1a,b,2a,b)}_1$
correspond to formulas ($\ref{F21a}$--$\ref{F12ab}$).

The absolute value of $\delta$ does not exceed 3
and it is much less than $\bar{l_6}$ and $\bar{l_5}$ but the
corrections due to
$\bar{l_5}$ and  $\bar{l_6}$ have a different sign and almost equal
values and therefore the account of $\delta$ is necessary.

\section{Conclusion}

$\quad$ Collecting all terms we can write final results for structure
functions $F_2$:
\be
\label{F2f}
F_2=\left(1-\frac{R_6 Q^2}{48\pi^2F_\pi^2}\right)
[F^{(1a)}_2 +F^{(1b)}_2+F^{(2a,b)}_2 /2]
+\frac{\alpha Q^2(\bar l_6 - \bar l_5+\delta)}{96\pi^3F_\pi^2 }
\left(\phi x -\frac{M^2}{\nu}L\right)
\ee
and
$$
F_1=\left(1-\frac{R_6 Q^2}{48\pi^2F_\pi^2}\right)
[F^{(1a)}_1 +F^{(1b)}_1 +F^{(2a,b)}_1 /2] +F^{(2c)}+
$$
\be
\label{F1f}
+\frac{\alpha Q^2[(\bar l_6 - \bar l_5+\delta)-R_6 x]}{192\pi^3F_\pi^2 x}
\left(\phi x -\frac{M^2}{\nu}L\right),
\ee
It is worth to note that we did not make any assumption about ratio
$M^2/Q^2$, so these results can be used from $Q^2=0$.

At $Q^2\approx 0.1-0.15$GeV$^2$ the large first order ChPT correction comes
from the factors in the square brackets in front of the first terms
in the r.h.s of ($\ref{F2f}$) and ($\ref{F1f}$) . These factors
have the meaning of the squares of pion form factors in the vertices
of the diagrams Figs. 1,2. Therefore, the accuracy of
($\ref{F2f}$) and $\ref{F1f}$) may be improved, if these
factors would be represented in the standard form of pion form factors
$(1+R_6Q^2/96\pi^2 F_\pi^2)^{-2}$. In the numerical calculations
we use such a procedure.

 The numerical results of the calculation at few values of
parameters are represented in Fig.~5-9. In the Fig. 5,6 one
can see the $Q^2$-dependence for $F_2$ and $F_1$ functions, respectively.
\begin{figure}
\centerline{\epsfxsize=12cm, \epsfbox{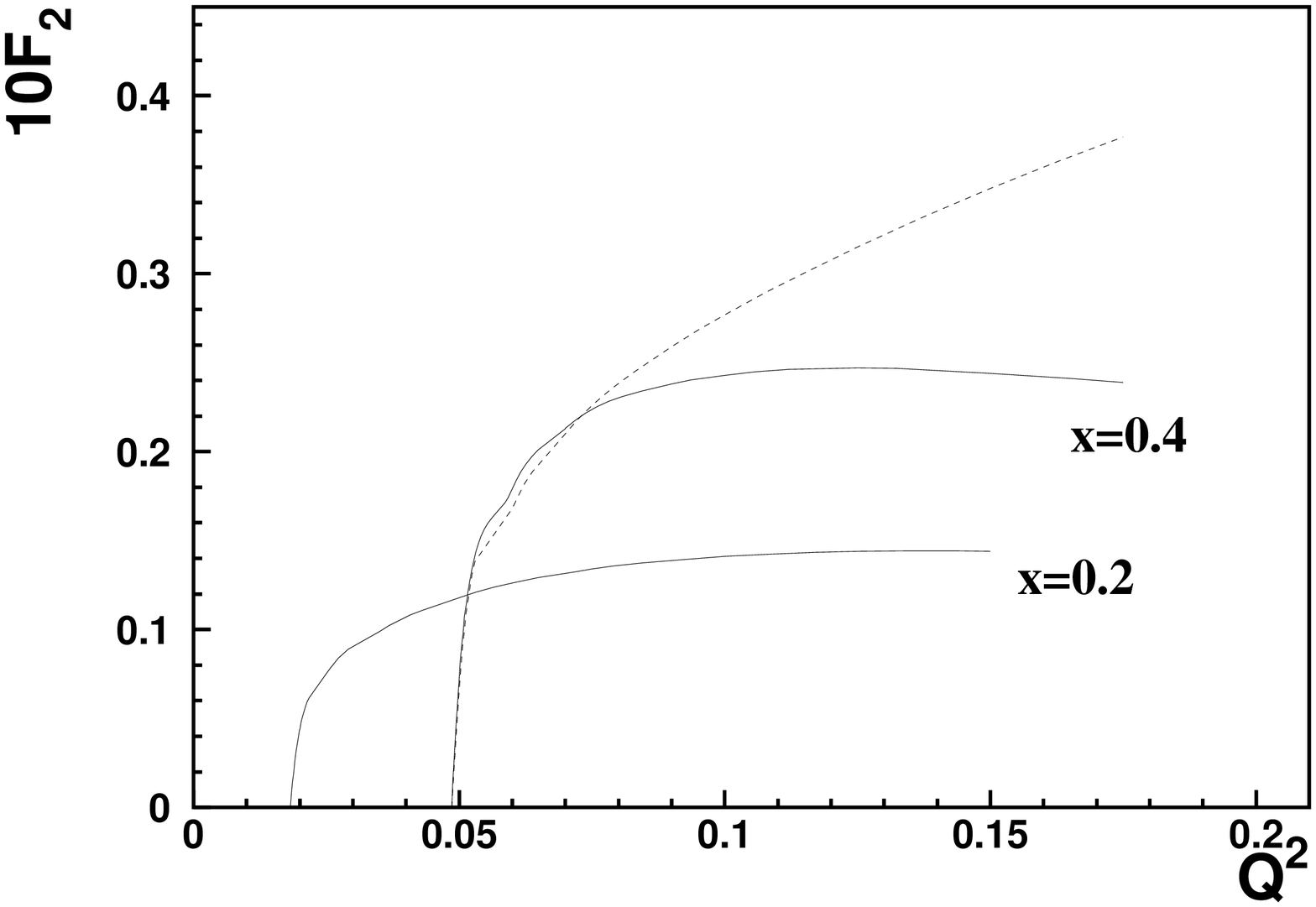}}
\caption{
The structure function $F_2$ as a function of $Q^2$
at fixed $x=0.2$ and $x=0.4$.
 The dashed line represents the
contribution of the scalar electrodynamics(Born)
 term at $x=0.4$.}
\end{figure}

\begin{figure}
\centerline{\epsfxsize=12cm, \epsfbox{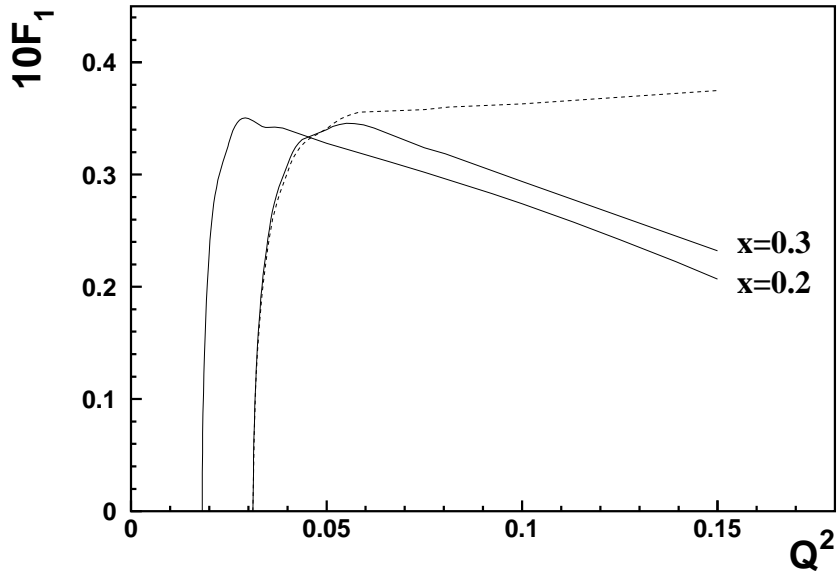}}
\caption{
The structure function $F_1$ as a function of $Q^2$
at fixed $x=0.2$ and $x=0.3$.
 The dashed line represents the
contribution of the scalar electrodynamics(Born)
 term at $x=0.3$.}
\end{figure}
We see that the zero order (Born term) and the chiral corrections to $F_2$
grows with the same rate and as a result $F_2$ becomes an almost constant.
For the case of $F_1$ the situations is different: the chiral corrections
grows faster then zero order term and $F_1$ decreases from some value of
$Q^2$. It is important to say that this behavior appear when chiral
corrections are rather small and ChPT is valid. In the scaling region of
high $Q^2$ numerical value of $F_1$ is approximately a few times as large
as $F_1$ at low $Q^2$ and from decreasing of $F_1$ it follows that  $F_1$
is not monotonous function of $Q^2$ but has a minimum at some intermediate $Q^2$.

The x-dependence of $F_2$ and $F_1$ structure function are
represented in Fig. 7,8, respectively. At large $x$ the structure functions vanishes due to
phase volume factor.
\begin{figure}
\centerline{\epsfxsize=12cm, \epsfbox{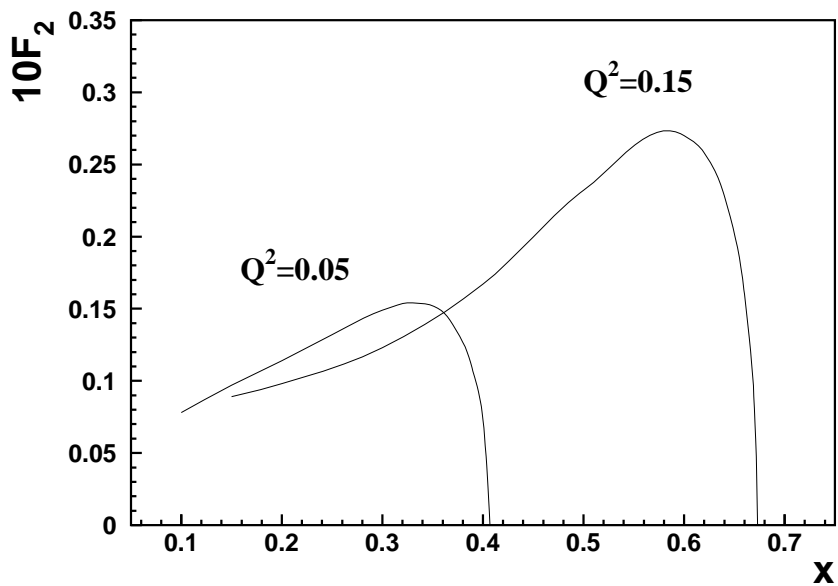}}
\caption{
The structure function $F_2$ as a function of $x$
at fixed $Q^2=0.05$GeV$^2$ and $Q^2=0.15$GeV$^2$.}
\end{figure}

\begin{figure}
\centerline{\epsfxsize=12cm, \epsfbox{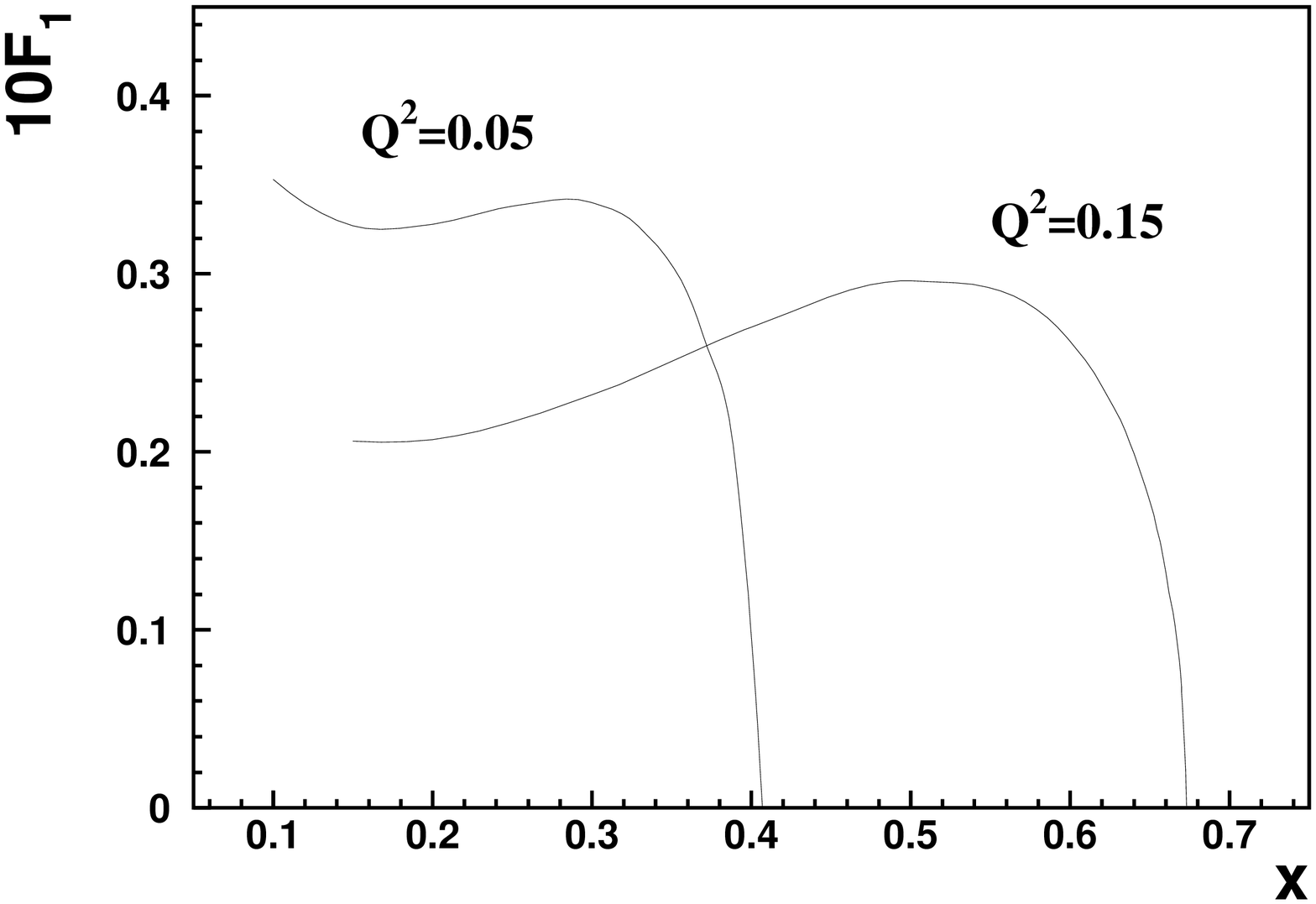}} \caption{ The
structure function $F_1$ as a function of $x$ at fixed
$Q^2=0.05$GeV$^2$ and $Q^2=0.15$GeV$^2$.}
\end{figure}

 Let us now discuss the accuracy of the obtained results.
Since only two terms in ChPT were calculated, only the general arguments,
referring to the convergence of ChPT can be used. According to these
 arguments the expansion parameters are the ratios
of all invariants entering in the problem - $Q^2$, and $s$
to the characteristic ChPT scale $\Lambda^2\simeq0.6$GeV$^2$. In our case
at $Q^2\simeq0.1-0.15$GeV$^2$ and this ratio
is of order $1/3$.
The total energy s enters only in the correction $\delta$
due to four pion interaction and at $x\ga0.15$
does not contribute to the structure functions more than $20\%$.
So, one may expect that the accuracy of obtained results is about
$30-50\%$. The accuracy is better at intermediate $x\sim0.2-0.3$ and worse
at $x\approx 0.15$ because the ratio $s/\Lambda^2$ becomes large.
It is not, however, a solid statement. From \cite{mark} we can see that ChPT
predictions for the cross section for
$\gamma\gamma\rightarrow\pi^+ \pi^-$ process are in a good agreement with
experiment up to $s\sim1$GeV$^2$.
The comparison of the obtained results with experiment would be
important and useful for understanding
the applicability domain of chiral theory,
the role of higher order ChPT corrections, etc.

\vspace{1.cm}

\centerline{\bf Acknowledgements}
This work was done under the partial support of CRDF grant RP2-2247,
INTAS grant 2000-587 and RFBR  grant 00-02-17808.

\end{document}